\documentstyle[preprint,aps,psfig]{revtex}
\tighten
\begin{document}
\preprint{\vbox{Submitted to Physical Review C}} 

\title{Self-consistent description of nuclear compressional modes}
\author{J. Piekarewicz}
\address{Department of Physics,
         Florida State University, 
         Tallahassee, FL 32306}
\date{\today}
\maketitle
\begin{abstract}
Isoscalar monopole and dipole compressional modes are computed for a
variety of closed-shell nuclei in a relativistic random-phase
approximation to three different parametrizations of the Walecka model
with scalar self-interactions. Particular emphasis is placed on the
role of self-consistency which by itself, and with little else,
guarantees the decoupling of the spurious isoscalar-dipole strength
from the physical response and the conservation of the vector
current. A powerful new relation is introduced to quantify the
violation of the vector current in terms of various ground-state
form-factors.  For the isoscalar-dipole mode two distinct regions are
clearly identified: (i) a high-energy component that is sensitive to
the size of the nucleus and scales with the compressibility of the
model and (ii) a low-energy component that is insensitivity to the 
nuclear compressibility. A fairly good description of both 
compressional modes is obtained by using a ``soft'' parametrization 
having a compression modulus of $K\!=\!224$~MeV.
\end{abstract}
\vspace{20pt}
\pacs{PACS numbers(s): 24.10.Jv, 21.10Re, 21.60.Jz}
\vfill
%
\section{Introduction}
\label{introduction}
The study of nuclear compressional modes, while interesting in its own
right, is motivated by our desire to understand the equation of state
of hadronic matter, especially in relation to its compression
modulus. In turn, an accurate determination of the compression modulus
places important constraints on theoretical models of nuclear
structure, heavy-ion collisions, neutron stars, and supernovae
explosions.

While it remains true that measuring the energy of the nuclear
compressional modes provides the most accurate determination of the
compression modulus, significant advances in astronomical observations
and terrestrial experiments are providing important complimentary
information. For example, explaining the time structure of the
neutrino burst emitted from supernova SN1987A seems to require a
relatively soft equation of state as input in the simulations of
core-collapsed supernova~\cite{Ma93,Wi93}. Further, the recently
inferred narrow mass distribution of neutron stars~\cite{TC99} poses
stringent constraints on the nuclear equation of state. At the same
time, a number of improved radii-measurements of radio-quite, 
isolated neutron stars --- such as RX J185635-3754 --- will contribute
significantly to our understanding of the high-density component of
the equation of state~\cite{Wa97}. Finally, measurements of the
elliptical flow in relativistic heavy-ion reactions seem to have
established the utility of this observable as a probe of the
stiffness of the equation of state~\cite{Da98}.

Also significant is the strong correlation between seemingly 
unrelated experiments. Indeed, the radius of a neutron star 
is predicted to be strongly correlated to the neutron skin of 
a heavy nucleus~\cite{Br00,Ho00}. Thus, the upcoming measurement 
of the neutron radius of ${}^{208}$Pb at the Jefferson 
Laboratory~\cite{Ho01,Mi00} should place important limits
on the radii of neutron-stars.

Although measurements of the giant monopole resonance~\cite{Yo77,Yo81}
and the isoscalar giant dipole resonance~\cite{Mo83,Ad86,Po92} have
existed for some time, the field has seen a revitalization due to new
and improved measurements of both compressional
modes~\cite{Da97,Cl99,Yo99}.  The field has also seen significant
advances in the theoretical domain. Indeed, calculations of nuclear
compressional modes using Hartree-Fock (HF) plus RPA approaches with
state-of-the-art Skyrme interactions are now
possible~\cite{Co99,Co00}. Relativistic RPA models have also enjoyed a
great deal of success, especially now that scalar self-interactions
have been incorporated into the calculation of the
response~\cite{Ma97,Ma99,Pi00,Vr00}. At the same time the philosophy
behind the theoretical extraction of the nuclear compressibility has
evolved considerably. Earlier attempts depended heavily on
semi-empirical formulas that related the compressibility to the
energies of the compressional modes~\cite{Bl99}. The field now demands
stricter standards: the model, without any recourse to semi-empirical
mass formulas, must predict both the compressibility of nuclear matter
as well as the energy of the compressional modes.

In this publication state-of-the-art calculations of the isoscalar
giant-monopole resonance (GMR) and the isoscalar giant-dipole
resonance (ISGDR) are reported for a variety of closed-shell
nuclei. This paper represents an expanded version of a short article
published recently that focused exclusively on
${}^{208}$Pb~\cite{Pi00}. The model adopted in this work is based on a
relativistic random-phase-approximation (RPA) to three different
parameterizations of the Walecka model with scalar self-interactions. 
A nonspectral approach that treats discrete and continuum excitations
on equal footing is implemented. As a result, the
conservation of the vector current is strictly maintained throughout
the calculation. Moreover, for the calculation of the RPA response we
employ a residual particle-hole interaction consistent with the
particle-particle interaction used to generate the mean-field ground
state. In this way the spurious isoscalar-dipole strength,
associated with the uniform translation of the center-of-mass, 
gets shifted to zero excitation energy and is cleanly separated 
from the physical response.

Having established the theoretical underpinning of our calculation, it
is now useful to contrast it against alternative self-consistent
implementations. In a recent article by Shlomo and
Sanzhur~\cite{Sh00}, it is suggested that actual implementations of
the RPA, in spite of claiming otherwise, are not fully
self-consistent. It is pointed out that these calculations often
resort to a variety of approximations such as: (i) neglecting the
two-body Coulomb and spin-orbit terms in the residual
particle-hole interaction, (ii) approximating the momentum-dependent
parts in the particle-hole interaction, (iii) limiting the
particle-hole space in a discretized calculation by a cut-off energy
$E_{ph}^{\rm max}$, and (iv) introducing a smearing parameter, such as
a Lorentzian width. Each of these approximations is now briefly
addressed. In the relativistic formalism employed here neither the
two-body Coulomb nor the spin-orbit interaction are neglected.
Rather, the residual particle-hole interaction includes the
(isoscalar) contribution from the photon as well as spin-orbit effects
that are incorporated --- to all orders --- by merely maintaining the
relativistic structure of the interaction. Moreover, the residual
particle-hole interaction is momentum independent because one preserves
intact its full Lorentz structure; no momentum-dependence is generated
through a nonrelativistic reduction of the interaction. Further, the
non-spectral approach employed here avoids any reliance on artificial
cutoffs and truncations. Finally, while a Lorentzian width is included
to compute the properties of discrete excitations, it is done so by
ensuring that the physically relevant quantities, the excitation
energy and the inelastic form-factor, remain invariant under a change
in width.

The paper has been organized as follows. Section~\ref{formalism}
describes the relativistic mean-field plus RPA formalism in great
detail placing special emphasis on the role of self-consistency.
Section~\ref{Fundamental Symmetries} illustrates the importance of
self-consistency for the conservation of the vector current and for
the decoupling of spurious strength from the physical isoscalar-dipole
response. Here a powerful novel relation is introduced to quantify the
violation of the vector current in terms of various known ground-state
form-factors. Results are displayed in Sec.~\ref{Results}, while a
summary and conclusions are presented in Sec.~\ref{Conclusions}.

\section{Formalism}
\label{formalism}
In this section a detailed description of the mean-field plus RPA 
formalism employed to compute the distribution of strength for both 
compressional modes is presented. This formalism, with the exception 
of its implementation in the case of scalar self-interactions, has now 
been available for almost fifteen years~\cite{Ni86,We88,Ho89}. However,
important lessons keep being ignored~\cite{Ma97}, just to be soon 
rediscovered~\cite{Ma99}. Thus, we feel compelled to present, for what 
we hope is the last time, a thorough discussion of the relativistic
RPA formalism.

The first step in calculating a relativistic RPA response is the 
computation of the mean-field ground state in a self-consistent 
approximation. Once self-consistency is achieved, three important 
pieces of information become available: (i) the single-particle 
energies of the occupied orbitals, (ii) their single-particle 
wave functions, and (iii) the self-consistent mean-field potential. 
This mean-field potential, without any modification, must then be 
used to generate the nucleon propagator; in this way the conservation 
of the vector current is guaranteed to be maintained. The nucleon 
propagator is computed nonspectrally to avoid any dependence on 
the artificial cutoffs and truncations that plague most spectral 
approaches. Moreover, through a nonspectral approach one gives
equal treatment to both bound and continuum orbitals.

Having generated the occupied single-particle spectrum and the nucleon 
propagator, the computation of the lowest-order (Hartree) polarization 
is reduced to the evaluation of various matrix elements of the
relevant transition operator.  To compute the RPA response one needs 
to go beyond the single-particle response. The RPA builds coherence 
among the many allowed particle-hole excitations by iterating the
lowest-order polarization to all orders via the residual
particle-hole interaction. Yet special care must be taken in adopting 
a residual particle-hole interaction consistent with the
particle-particle interaction used to generate the mean-field ground 
state. Only then can one ensure that the spurious component of the 
isoscalar-dipole response will get shifted to zero excitation 
energy~\cite{Th61,DF90}. As the polarization tensor is a fundamental 
many-body operator, it can be computed systematically using well-known 
many-body techniques~\cite{FW71}. Having computed the polarization
tensor, the nuclear response is extracted by simply taking its 
imaginary part. The following sections provide a detailed account on 
the implementation of these ideas.

\subsection{The Lagrangian Density}
\label{lagrangian}

The starting point for the calculation of the nuclear response
is a Lagrangian density having an isodoublet nucleon field 
($\psi$) interacting via the exchange of two isoscalar mesons, 
the scalar sigma ($\phi$) and the vector omega ($V^{\mu}$), 
one isovector meson, the rho ($b^{\mu}$), and the photon 
($A^{\mu}$)~\cite{SW86,SW97}. The pseudoscalar pion is not 
included as it does not contribute at the mean-field level. 
In addition to meson-nucleon interactions the Lagrangian density 
includes scalar self-interactions. These are responsible for reducing 
the nuclear compressibility from the unrealistically large value of 
$K\!=\!545$~MeV, obtained in the original linear model of 
Walecka~\cite{Wa74}, all the way down to the acceptable value 
of $K\!=\!224$~MeV. Thus, without the inclusion of scalar 
self-interactions a realistic calculation of the compressional 
modes is not feasible. The Lagrangian density for the model is 
thus given by
\begin{equation}
 {\cal L}_{\rm int} = \bar{\psi}\Big[
      g_{s}\phi - 
      g_{\rm v}\rlap\slash V -
      {g_{\rm \rho}\over2}{\mbox{\boldmath$\tau$}}
      \cdot\rlap\slash{\bf b} -
      {e \over 2}(1+\tau_{3})\,\rlap\slash{\!A} \Big]\psi - 
      U(\phi) \;; \quad
      U(\phi) = \frac{1}{3!}\kappa\phi^{3}
              + \frac{1}{4!}\lambda\phi^{4} \;,
\end{equation}
were use of the``slash'' notation,
$\rlap\slash V \!\equiv \gamma^{\mu}V_{\mu}$,
has been made. The various model parameters have been
listed in Table~\ref{Table1}.
\subsection{The nucleon propagator}
\label{nucprop}
\begin{figure}[h]
\vskip0.25in
\leavevmode\centering\psfig{file=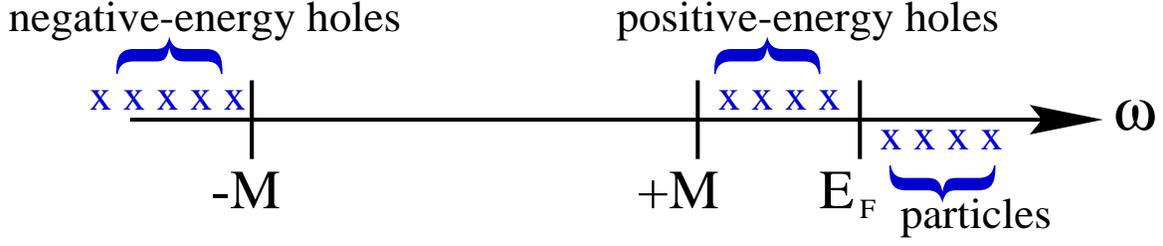,width=6in}
 \caption{Spectral content of the nucleon propagator
          in a relativistic Fermi-gas approximation.}
 \label{Figure1}
\end{figure}

The mean-field propagator contains information about the interaction 
of the propagating nucleon with the average potential generated by the 
nuclear medium. However, even in a Fermi-gas description, where all 
interactions are neglected, the nucleon propagator would still differ 
from its free-space value because of the presence of a filled Fermi 
sea. Indeed, the analytic structure of the free-nucleon propagator at 
finite density is different from its free-space value 
(see Fig.~\ref{Figure1}). This suggests the following decomposition 
of the nucleon propagator~\cite{SW86}:
\begin{mathletters}
\begin{eqnarray}
  &&
  G(x,y) = \int_{-\infty}^{\infty} {d\omega \over 2\pi}
           e^{-i\omega(x^{0}-y^{0})}
           G({\bf x},{\bf y};\omega) \;, \\
  &&
    G({\bf x},{\bf y};\omega)     =
    G_{F}({\bf x},{\bf y};\omega) +
    G_{D}({\bf x},{\bf y};\omega) \;.
 \label{gxy}
\end{eqnarray}
\end{mathletters}
The Feynman part of the propagator, $G_{F}$, admits a spectral 
decomposition in terms of the mean-field solutions to the Dirac 
equation. That is, 
\begin{equation}
  G_{F}({\bf x},{\bf y};\omega) = \sum_{n}
   \left[
     \frac{U_{n}({\bf x})\overline{U}_{n}({\bf y})}
          {\omega - E_{n}^{(+)} + i\eta} + 
     \frac{V_{n}({\bf x})\overline{V}_{n}({\bf y})}
          {\omega + E_{n}^{(-)} - i\eta}  
   \right] \;,
 \label{gfeyn}
\end{equation}
where $U_{n}$ and $V_{n}$ are the positive- and negative-energy
solutions to the Dirac equation, and the sum is over all states 
in the spectrum. The analytic structure of $G_{F}$ is identical
to that of the conventional Feynman propagator~\cite{BD65}. The 
density-dependent part of the propagator, $G_{D}$, corrects 
$G_{F}$ for the presence of a filled Fermi sea. This correction 
occurs even in a noninteracting system and is due to the Pauli 
exclusion principle. Formally, one effects this correction by
shifting the position of the pole of every occupied state from 
below to above the real axis (see Fig.~\ref{Figure1})
\begin{eqnarray}
    G_{D}({\bf x},{\bf y};\omega) &=& \sum_{n < {\rm F}}
     U_{n}({\bf x})\overline{U}_{n}({\bf y}) 
     \left[
        {1 \over \omega - E_{n}^{(+)} - i\eta} -
        {1 \over \omega - E_{n}^{(+)} + i\eta} 
     \right] \nonumber \\ &=&
     2 \pi i \sum_{n < {\rm F}}
     \delta\Big(\omega-E_{n}^{(+)}\Big)     
               {U}_{n}({\bf x})        
      \overline{U}_{n}({\bf y})        \;.
 \label{gdens}
\end{eqnarray}
Note that the sum over $n$ is now restricted to only those
positive-energy states below the Fermi energy. In a mean-field
approximation these states satisfy a Dirac equation of the form:
\begin{equation}
  \left[ 
   E_{n}^{(+)}\gamma^{0}+i{\bf\gamma}\cdot{\bf \nabla}
                        -M-\Sigma_{\rm MF}(x)
  \right]U_{n}({\bf x})=0 \;,
 \label{Uspinor}
\end{equation}
where the mean-field potential is given by
\begin{equation}
 \Sigma_{\rm MF}(x) = \Sigma_{\rm S}(x) 
                    + \gamma^{0}\Sigma_{0}(x) \;.
\label{Sigma}
\end{equation}
The quantities $\Sigma_{\rm S}$ and $\Sigma_{0}$ denote the scalar 
and vector potentials that have been generated self-consistently 
at the mean-field level. Since this work is limited to the response 
of closed-shell nuclei, it is assumed that the mean-field potential 
has been generated by a spherically-symmetric, spin-saturated ground 
state. 

Although the above spectral decomposition of the nucleon propagator 
will become important in understanding the spectral content of the 
nuclear response, in practice it suffers from a reliance on artificial 
cutoffs and truncations. An efficient scheme that avoids such a 
dependence is the nonspectral approach. A nonspectral approach has 
the added advantage that both positive- and negative-energy continuua 
are treated exactly. As a result, the contributions from the 
negative-energy states to the response are included automatically. 
This is important to maintain fundamental physical principles, as 
the positive-energy states by themselves are not complete. To obtain 
the nucleon propagator in nonspectral form one must solve the
following inhomogeneous Dirac equation:
\begin{equation}
  \left[ 
   \omega\gamma^{0}+i{\bf\gamma}\cdot{\bf \nabla}-M-\Sigma_{\rm MF}(x)
  \right]G_{F}({\bf x},{\bf y};\omega)=
  \delta({\bf x}-{\bf y}) \;.
 \label{DiracEq}
\end{equation}
Here $\omega$ is taken to be a complex variable and the mean-field 
potential is identical to the one used to generate the nuclear ground 
state. Taking advantage of the spherical symmetry of the potential, 
one may decompose the Feynman propagator in terms of spin-spherical 
harmonics
\begin{equation}
  G_{F}({\bf x},{\bf y};\omega)=\frac{1}{xy}\sum_{\kappa m}
   \left(
   \matrix{
           \phantom{i}
           g_{11}^{\kappa}(x,y;\omega)
           \langle\hat{\bf x}|\!+\!\kappa m\rangle
           \langle +\kappa m |\hat{\bf y}\rangle &
         -ig_{12}^{\kappa}(x,y;\omega)
           \langle\hat{\bf x}|\!+\!\kappa m\rangle
           \langle -\kappa m |\hat{\bf y}\rangle \cr
          ig_{21}^{\kappa}(x,y;\omega)
           \langle\hat{\bf x}|\!-\!\kappa m\rangle
           \langle +\kappa m |\hat{\bf y}\rangle &
           \phantom{-i}
           g_{22}^{\kappa}(x,y;\omega)
           \langle\hat{\bf x}|\!-\!\kappa m\rangle
           \langle -\kappa m |\hat{\bf y}\rangle }
   \right) \;,
 \label{gij}
\end{equation}
which are defined as
\begin{eqnarray}
  &&
  \langle\hat{\bf x}|\kappa m\rangle  =
  \sum_{m_{l}m_{s}}
  \langle lm_{l},{\scriptstyle{\frac{1}{2}}}m_{s}|
  l{\scriptstyle{\frac{1}{2}}}jm\rangle 
  Y_{lm_{l}}(\hat{\bf x})\chi_{{\scriptstyle{\frac{1}{2}}}m_{s}}\;,
 \label{curlyy} \\
  && \,
  j\!=\!|\kappa|\!-\!\frac{1}{2}\; \quad\hbox{and}\quad
  l=\cases{+\kappa   & if $\kappa>0\;,$ \cr
           -\kappa-1 & if $\kappa<0\;.$   }
 \label{ljkappa}
\end{eqnarray}
The above decomposition enables one to rewrite the Dirac equation
as a set of first-order, coupled, ordinary differential equations 
of the form
\begin{equation}
  \left(
   \matrix{
   \omega^{*}\!-\!M^{*} & 
   \displaystyle{\frac{d}{dx}\!-\!\frac{\kappa^{*}}{x}} \cr
   \displaystyle{\frac{d}{dx}\!+\!\frac{\kappa^{*}}{x}} &
  -\omega^{*}\!-\!M^{*} }
  \right)
  \left(
   \matrix{g_{11}^{\kappa} & g_{12}^{\kappa} \cr
                           &                 \cr
           g_{21}^{\kappa} & g_{22}^{\kappa} }
  \right)=\delta(x-y) \;,
 \label{DiracEqG}
\end{equation}
where we have defined
\begin{equation}
  \omega^{*}\equiv \omega-\Sigma_{\rm v}(x) 
  \;\hbox{and}\;\;\;
  M^{*}\equiv M+\Sigma_{\rm s}(x)\;.
 \label{stars}
\end{equation}
It is important to underscore that the mean-field potentials used 
to compute the nucleon propagator must be identical to those used 
to generate the mean-field ground state if the conservation of the 
vector current is to be maintained.
\subsection{The nuclear polarization}
\label{nucpolar}

To illustrate the many-body techniques employed in the manuscript,
we define a general polarization insertion as the time-ordered
product of two arbitrary nucleon currents:
\begin{equation}
  i\Pi^{\alpha\beta}(x,y) =  \langle \Psi_{0}| 
  T \Big[ \hat{J}^{\alpha}(x)\hat{J}^{\beta}(y)\Big] |\Psi_{0}\rangle \;,
 \label{pialphabeta}
\end{equation}
where $\Psi_{0}$ denotes the exact nuclear ground state and 
$\hat{J}^{\alpha}(x)$ is a one-body current operator of the form
\begin{equation}
  \hat{J}^{\alpha}(x) = \bar{\psi}(x)\Gamma^{\alpha}\psi(x) \;.
 \label{jalpha}
\end{equation}
Note that the ``big'' gamma matrices have been defined so that 
the one-body current operator be hermitian~\cite{BD65}. That 
is,
\begin{equation}
\Gamma^{\alpha}=\{1,i\gamma^{5},\gamma^{\mu},
             \gamma^{\mu}\gamma^{5},\sigma^{\mu\nu}\}
             \quad {\rm with} \quad
             \overline{\Gamma}^{\alpha}
             \equiv
             \gamma^{0}\,\Gamma^{\alpha\dagger}\,\gamma^{0} 
           = \Gamma^{\alpha}\;.
\label{BigGammas}
\end{equation}
In a mean-field approximation to the nuclear ground state, such as 
the one employed here and in most of the other relativistic 
calculations to date, the polarization insertion may be written 
exclusively in terms of the nucleon mean-field propagator
\begin{equation}
  i\Pi^{\alpha\beta}(x,y) = {\rm Tr} 
  \Big[\Gamma^{\alpha}G(x,y)\Gamma^{\beta}G(y,x)\Big] \;.
 \label{pialphabetamf}
\end{equation}

The earlier decomposition of the nucleon propagator into Feynman 
and density-dependent contributions [Eq.~(\ref{gxy})] suggests an 
equivalent decomposition for the polarization insertion
\begin{mathletters}
\begin{eqnarray}
  &&
  \Pi^{\alpha\beta}(x,y) = \int_{-\infty}^{\infty} {d\omega \over 2\pi}
              e^{-i\omega(x^{0}-y^{0})}
             \Pi^{\alpha\beta}({\bf x},{\bf y};\omega) \;, \\
  &&
  \Pi^{\alpha\beta}({\bf x},{\bf y};\omega)     =
  \Pi^{\alpha\beta}_{F}({\bf x},{\bf y};\omega) +
  \Pi^{\alpha\beta}_{D}({\bf x},{\bf y};\omega) \;.
 \label{pixyfd}
\end{eqnarray}
\end{mathletters}
The Feynman part of the polarization, $\Pi^{\alpha\beta}_{F}$, is
independent of $G_{D}$ and describes the polarization of the vacuum.
This piece, which diverges and needs to be renormalized, has been
incorporated in our earlier calculations of the longitudinal response
in the quasifree region~\cite{HP89}. However, it has been included
only in a local-density approximation. To our knowledge an exact
finite-nucleus calculation of vacuum polarization has yet to be
performed. While a local-density approximation is accurate in the
quasifree region where many angular-momentum channels contribute, it
has proven inadequate for the description of discrete nuclear
excitations~\cite{Pi90}. In particular, the spurious isoscalar dipole
strength associated with the uniform translation of the center of mass
does not get shifted all the way down to zero excitation energy. More
relevant, the role of vacuum polarization in effective hadronic
theories is currently being revisited. Effective Field Theories now
suggest that the largely unknown physics associated with the
short-distance dynamics may be effectively simulated by the use of
various local operators~\cite{Le97,Be99,Fu00}. It is for these reasons
that vacuum polarization will be ignored henceforth.  Note, however,
that it is still possible to ignore vacuum effects and end up with a
completely consistent model of the nuclear response~\cite{DF90,SW86}.

In contrast to the Feynman part of the polarization, the  
density-dependent part is finite and can be computed exactly in the 
finite system~\cite{Ni86,We88,Ho89}. It is given by
\begin{equation}
  \Pi_{D}^{\alpha\beta}({\bf x},{\bf y};\omega)  \equiv
  \Pi_{FD}^{\alpha\beta}({\bf x},{\bf y};\omega) +
  \Pi_{DF}^{\alpha\beta}({\bf x},{\bf y};\omega) \;,
 \label{pidens}
\end{equation}
where
\begin{mathletters}
\begin{eqnarray}
  \Pi_{FD}^{\alpha\beta}({\bf x},{\bf y};\omega) &=&
   \sum_{n<F} 
   \overline{U}_{n}({\bf x})\Gamma^{\alpha}\,
    G_{F}\Big({\bf x},{\bf y};+\omega+E_{n}^{(+)}\Big) 
    \Gamma^{\beta}U_{n}({\bf y})  \;,  \label{pifd} \\
  \Pi_{DF}^{\alpha\beta}({\bf x},{\bf y};\omega) &=&
   \sum_{n<F} 
   \overline{U}_{n}({\bf y})\Gamma^{\beta}\,
    G_{F}\Big({\bf y},{\bf x};-\omega+E_{n}^{(+)}\Big) 
    \Gamma^{\alpha}U_{n}({\bf x})  \;. \label{pidf} 
\end{eqnarray}
\label{pifddf}
\end{mathletters}
Note that the Pauli-blocking of particle-hole excitations, a term 
usually denoted by $\Pi^{\alpha\beta}_{DD}$, has already been 
incorporated in the above two terms.
The density-dependent part of the polarization includes the 
excitation of particle-hole pairs plus the mixing between 
positive- and negative-energy states; this last term is
sometimes referred to as the Pauli blocking of $N\bar{N}$ 
excitations. The spectral content of $\Pi_{D}$ is easily 
revealed by using the spectral decomposition of the Feynman 
propagator [see Eq.~(\ref{gfeyn})]. For example, the 
Feynman-density component of the polarization, 
$\Pi_{FD}^{\alpha\beta}$, may be written as
\begin{equation}
  \Pi_{FD}^{\alpha\beta}({\bf x},{\bf y};\omega) \!=\!
   \sum_{m,n<F} 
   \left[
     {\overline{U}_{n}({\bf x})\Gamma^{\alpha} 
     U_{m}({\bf x})\overline{U}_{m}({\bf y})
     \Gamma^{\beta}U_{n}({\bf y}) \over
     \omega - \Big(E^{(+)}_{m}-E^{(+)}_{n}\Big) + i\eta} 
     \!+\!
     {\overline{U}_{n}({\bf x})\Gamma^{\alpha}
     V_{m}({\bf x})\overline{V}_{m}({\bf y})
     \Gamma^{\beta}U_{n}({\bf y}) \over
     \omega + \Big(E^{(-)}_{m}+E^{(+)}_{n}\Big) - i\eta} 
   \right] \;.
 \label{pidspect}
\end{equation}   
The first term in the sum represents the excitation of a particle-hole
pair. The excitation becomes real, namely both particles go on-shell,
when the energy transfer to the nucleus becomes identical to the
pair-excitation energy $\omega\equiv E^{(+)}_{m}-E^{(+)}_{n}$. The
second term in the sum has no nonrelativistic counterpart; it
represents the mixing between positive- and negative-energy states.
Although the contribution from vacuum polarization has been neglected,
the inclusion of this mixing is of utmost importance for maintaining
current conservation. Moreover, it is also essential for the removal 
of all spurious strength from the excitation of the isoscalar dipole
mode. The inclusion of the negative-energy sector in the calculation 
of the response underscores the basic fact that the positive-energy 
sector of the spectrum, by itself, is not complete.
\subsection{The RPA equations}
\label{rpa}
The polarization tensor describes modifications to the propagation of
various mesons (such as the $\sigma$, $\omega$, $\rho$, $\ldots$) as 
they move through the nuclear environment. In addition, the 
polarization tensor contains all information on the excitation 
spectrum of the nucleus. Indeed, the polarization insertion is an 
analytic function of the frequency $\omega$, except for the presence 
of simple poles located at the excitation energies of the system.
The residue at the pole is simply related to the inelastic 
form-factor~\cite{FW71}.

The singularity structure of the lowest-order polarization tensor is
easily inferred from the mean-field spectrum: the nuclear excitation
energies (poles) appear at energies given by the difference between
the single-particle energies of a nucleon above the Fermi level
(particle) and one below (hole). In this approximation the residual
interaction between the particle and the hole is neglected. However,
the consistent response of the mean-field ground state demands that
the residual interaction between the particle and the hole be
incorporated~\cite{DF90}. This may be implemented by solving Dyson's
equation for the polarization insertion in a random-phase
approximation. In RPA the lowest-order polarization is iterated to all
orders via the residual particle-hole interaction. Because the
iteration is to all orders, the singularity structure of the
propagator, and thus the location of the poles, is modified relative
to the lowest-order predictions. Dyson's equation for the RPA
polarization is given by:
\begin{equation}
  \Pi^{\alpha\beta}_{\rm RPA}({\bf q},{\bf q}';\omega) =
  \Pi^{\alpha\beta}_{D}({\bf q},{\bf q}';\omega) \!+\!
  \int\!\frac{d^3k}{(2\pi)^{3}}
       \frac{d^3k'}{(2\pi)^{3}}
  \Pi^{\alpha\lambda}_{D}({\bf q},{\bf k};\omega)           
   V_{\lambda\sigma}({\bf k},{\bf k}';\omega)
  \Pi^{\sigma\beta}_{\rm RPA}({\bf k}',{\bf q}';\omega) \;,
 \label{PiabRPA} 
\end{equation}
where $V_{\lambda\sigma}({\bf k},{\bf k}';\omega)$ is the 
residual interaction to be discussed below and 
$\Pi^{\alpha\beta}_{D}({\bf q},{\bf q}';\omega)$ is the Fourier
 transform of the lowest-order polarization. That is,
\begin{equation}
  \Pi^{\alpha\beta}_{D}({\bf q},{\bf q}';\omega) =
  \int{d^{3}x}\,{d^{3}y}\, 
  e^{-i({\bf q} \cdot{\bf x}-{\bf q}'\cdot{\bf y})}\,
  \Pi^{\alpha\beta}_{D}({\bf x},{\bf y};\omega) \;.
 \label{piqq}
\end{equation}
At this point it is convenient to depart from the general formalism
adopted until now and restrict the discussion to the case of interest:
the isoscalar compressional modes. Hence, the only component of the
residual interaction that must be retained is the one mediated by the
exchange of the sigma and omega mesons, and the (isoscalar component
of the) photon. Moreover we employ the simplest operator, the timelike
component of the vector current
\begin{equation}
 \hat{\rho}({\bf q}) = \int{d^{3}x}\,e^{i{\bf q}\cdot{\bf x}}\,
                       \bar{\psi}({\bf x})\gamma^{0}\psi({\bf x}) \;,
\label{timelike}
\end{equation}
that can couple to these natural-parity excitations.

The computational demands imposed on a calculation of the RPA 
response for a nucleus as large as ${}^{208}$Pb can be formidable 
indeed. Powerful symmetries that are present in infinite nuclear 
matter, such as translational invariance, are broken in the finite 
system. As a result, the RPA equations that were algebraic in the 
infinite system become integral equations in the finite nucleus. 
Moreover, modes of excitation that were uncoupled before, such 
as longitudinal and transverse modes, become coupled now. In 
this way the RPA equations, because of the ubiquitous
scalar-longitudinal mixing, become a complicated $5\!\times\!5$ 
set of coupled integral equations. Correspondingly, the residual 
particle-hole interaction, also a $5\!\times\!5$ kernel, may be
written as
\begin{equation}
 V_{\alpha\beta}({\bf k},{\bf k}';\omega) = 
  \pmatrix{ 
    g_{\rm s}^{2}\Delta({\bf k},{\bf k}';\omega) & 0 \cr
    0 & g_{\rm v}^{2}D_{\alpha\beta}({\bf k},{\bf k}';\omega)} \;,
 \label{ResInt}
\end{equation}
where the vector propagator is given by
\begin{equation}
 D_{\alpha\beta}({\bf k},{\bf k}';\omega) =
 (2\pi)^{3}\delta({\bf k}-{\bf k}') 
 \left(
  -g_{\alpha\beta}+\frac{k_{\alpha}k_{\beta}}{m_{\rm v}^{2}}
 \right)D({\bf k},\omega)\;; \quad 
 D({\bf k},\omega)=\frac{1}{\omega^{2}-{\bf k}^{2}-m_{\rm v}^{2}} \;.
 \label{vecprop}
\end{equation}
Note that because vector self-interactions have not yet been included
in the present version of the model, the vector propagator remains
local (in momentum space) and maintains its simple Yukawa form. In 
contrast, scalar self-interactions modify the propagator relative 
to its simple free-space form. Hence, the scalar propagator now 
satisfies a nontrivial Klein-Gordan equation of the 
form:~\cite{Ma97,Ma99} 
\begin{equation}
  \Big(\omega^{2}\!+\!\nabla^{2}\!-m_{s}^{2}\!-\!U''(\phi)\Big)
  \Delta({\bf x},{\bf y}; \omega)\!=\!\delta({\bf x}\!-\!{\bf y})\;.
 \label{KG}
\end{equation}
\section{Fundamental Symmetries}
\label{Fundamental Symmetries}
In the following two sections we discuss important symmetries related
to the conservation of the vector current and to the elimination of
the spurious isoscalar-dipole strength from the physical response. We
are adamant about the preservation of these two fundamental symmetries
of nature as we regard the predictions of theoretical formulations
that violate them as ambiguous at best. For example, in a framework 
that violates the conservation of the vector current should one 
calculate the longitudinal response of the nuclear ground state by 
using the timelike component or the longitudinal one? Likewise, the 
predicted energy and distribution of isoscalar-dipole strength in a 
model that retains even a small fraction of spurious strength will 
bear little resemblance to reality. It is only through consistency,
the recurring theme of this paper, that one can enforce these 
important dynamical demands. How is that consistency plays such 
an important role in preserving these fundamental symmetries,
will now be discussed.
\subsection{Conservation of the vector current}
\label{currcons}
\begin{figure}[h]
\vskip0.25in
\leavevmode\centering\psfig{file=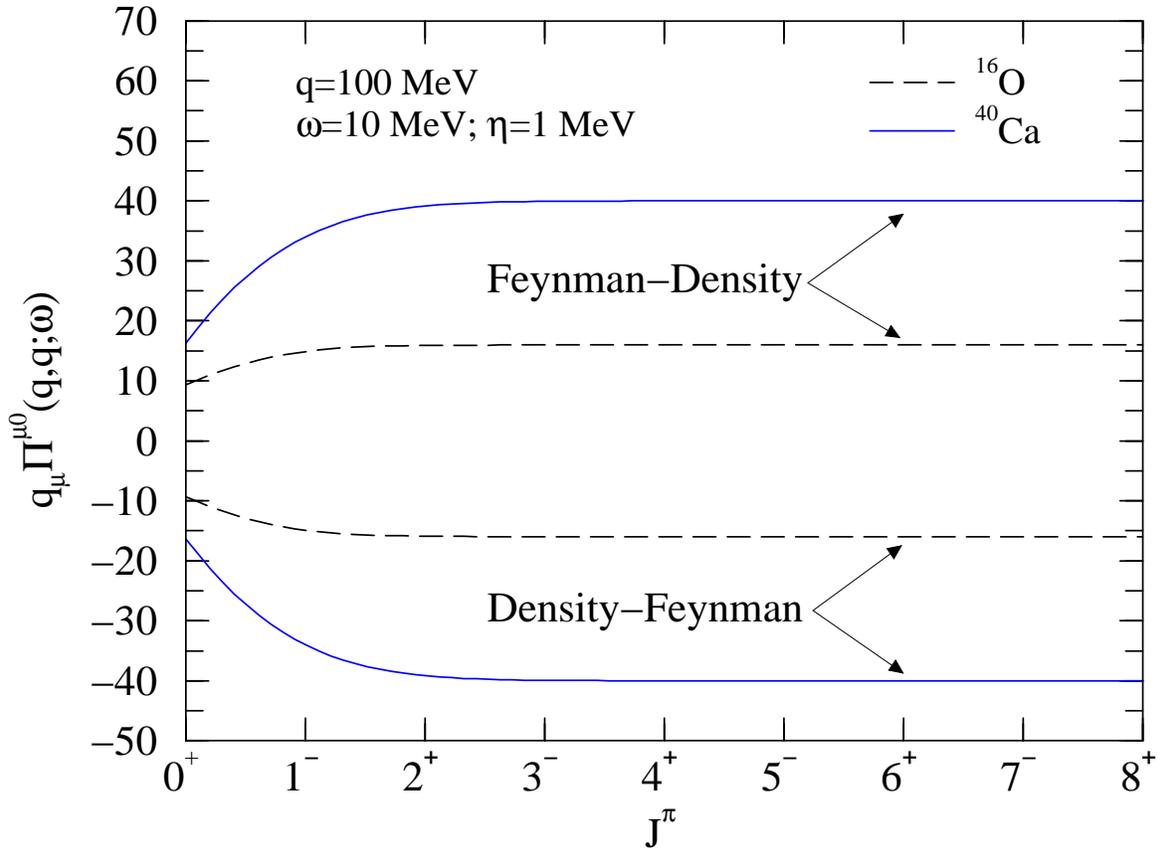,width=6in}
 \caption{The real part of $q_{\mu}\Pi^{\mu0}$ for the
          Feynman-Density and Density-Feynman parts of
          the nuclear polarization as a function of
	  the total angular momentum channel. Results 
	  are reported for ${}^{16}$O and ${}^{40}$Ca at
          $q\!=\!q'\!=\!100$~MeV and $\omega\!=\!10$~MeV. 
	  In a consistent mean-field formalism these 
	  quantities should approach $\pm A$, respectively.}
 \label{Figure2}
\end{figure}

\begin{figure}[h]
\vskip0.25in
\leavevmode\centering\psfig{file=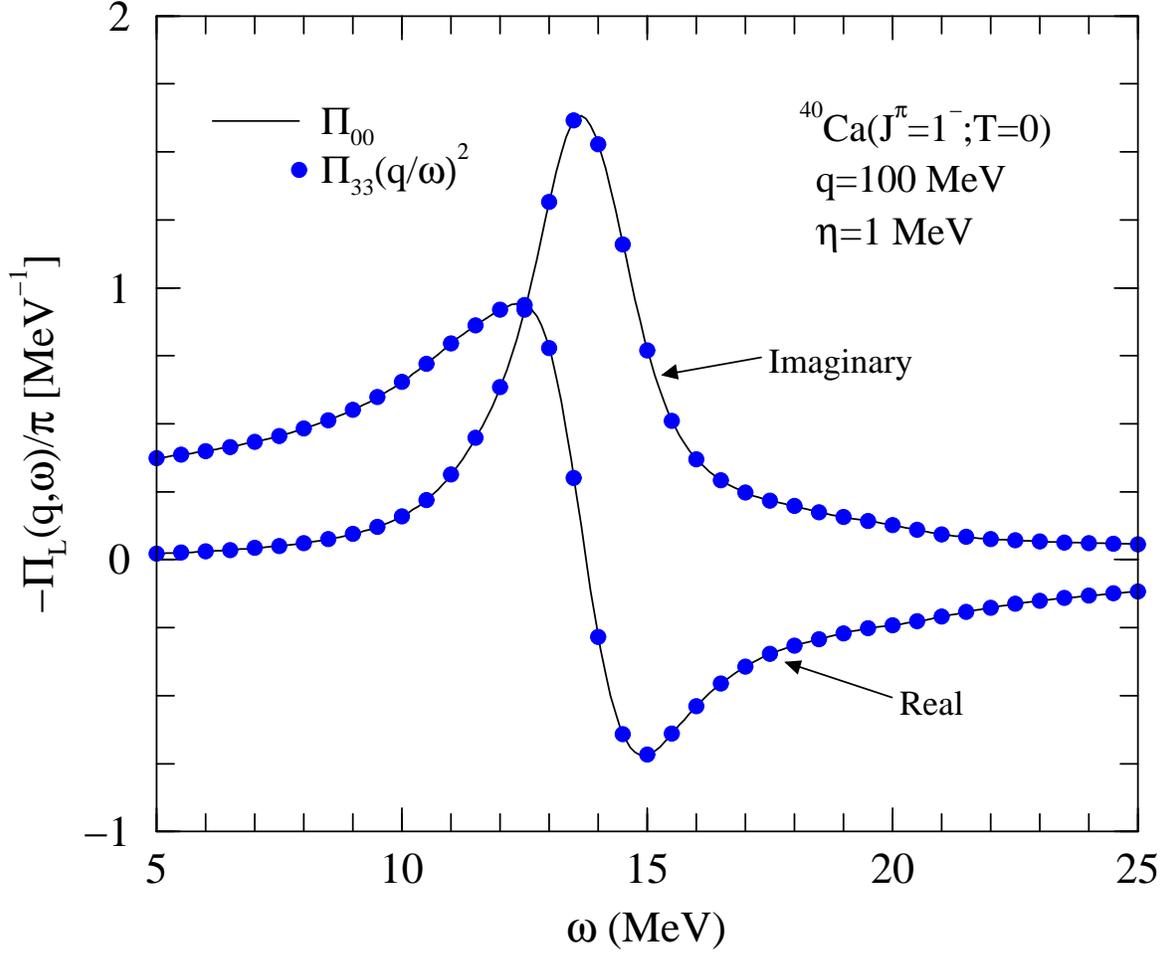,width=6in}
 \caption{The longitudinal polarization for the isoscalar
	  dipole state computed from the timelike component 
	  of the vector current (solid line) and from the 
	  longitudinal component (filled circles). In a 
	  consistent mean-field formalism --- such as the 
	  one used here --- they should be identical. Note 
	  that the imaginary component is the longitudinal 
	  response.}
 \label{Figure3}
\end{figure}

We start by discussing the conservation of the vector current.
Current conservation demands that the timelike component of
vector current be related to the longitudinal component. This 
impacts greatly on the results; it forces the nuclear polarization 
with one Lorentz vector index to be transverse to the 
four-momentum transfer, irrespective of the Lorentz character 
of the other vertex. That is,
\begin{equation}
  q_{\mu}\Pi^{\mu\beta}_{D}({\bf q},{\bf q}';\omega)=0 \quad
  {\rm with} \quad q^{\mu}=(\omega,{\bf q}) \;.
 \label{CC}
\end{equation}
So how is current conservation realized in our model? As indicated in
Eq.~(\ref{pidens}) the density dependent part of the polarization
tensor consists of two terms: $\Pi^{\alpha\beta}_{FD}$ and
$\Pi^{\alpha\beta}_{DF}$. 
Does each term separately satisfy current conservation or
does the conservation of the current depend on a sensitive
cancellation between them? To address this question we introduce the
longitudinal (with respect to ${\bf q}$) component of the vector
current. We start with the Feynman-density piece:
\begin{equation}
   q\Pi_{FD}^{3\beta}({\bf q},{\bf q}';\omega) \!=\!
   \int{d^{3}x}\,{d^{3}y}\, 
   \sum_{n<F} \overline{U}_{n}({\bf x})
   \big({\mbox{\boldmath$\gamma$}}\cdot{\bf q}\big)
    e^{-i{\bf q}\cdot{\bf x}}\,
    G_{F}\Big({\bf x},{\bf y};\omega+E_{n}^{(+)}\Big) 
    e^{i{\bf q}'\cdot{\bf y}}\,
    \Gamma^{\beta}U_{n}({\bf y}) \;.
 \label{qPiFD}
\end{equation}
To make contact with the timelike component of the polarization 
we turn the momentum transfer ${\bf q}$ into a gradient operator 
$
[\big({\mbox{\boldmath$\gamma$}}\cdot{\bf q}\big)
  e^{-i{\bf q}\cdot{\bf x}} \equiv 
  \big({\mbox{\boldmath$\gamma$}}\cdot i\nabla\big)
  e^{-i{\bf q}\cdot{\bf x}}] 
$
and integrate by parts. In this way the gradient operator acts now 
on both the bound-state nucleon spinor and the nucleon propagator.
It is then the difference between their respective Dirac equations 
[Eqs.~(\ref{Uspinor}) and~(\ref{DiracEq})] that dictates how severe 
the violation of current conservation becomes. We obtain
\begin{equation}
 q_{\mu}\Pi^{\mu\beta}_{FD}({\bf q},{\bf q}';\omega) = 
 \rho^{\beta}({\bf q}\!-\!{\bf q}') \equiv
 \int d^3x\,e^{-i({\bf q}\!-\!{\bf q}')\cdot{\bf x}}\,
 \sum_{n<{\rm F}}
 \overline{U}_{n}({\bf x})\Gamma^{\beta}{U}_{n}({\bf x}) \;,
 \label{CCPiFD}
\end{equation}
where $\rho^{\beta}({\bf q})$ represents a ground-state form-factor.
This is a new and important result. First, such a simple relation
would have been impossible to obtain had the mean-field potential for
the nucleon propagator been any different than the corresponding one
for the bound-state wave function. This is one of the many
manifestations of consistency in the formalism. Second, because in
spherical nuclei all form-factors are real~\cite{SW86}, the imaginary 
part of $\Pi^{\mu\beta}_{FD}$, by itself, satisfies current conservation. 
However, this is not true for the real part. Indeed, the violation 
to the real part of the polarization is regulated by the various 
ground-state form-factors. This result may be used as a stringent 
test on the numerics. For instance, if one lets
$\Gamma^{\beta} \rightarrow \gamma^{0}$ and sets 
${\bf q}\!=\!{\bf q}'$ in Eq.~(\ref{CCPiFD}), the 
violation becomes identical to the mass number of the 
nucleus. That is,
\begin{equation}
 q_{\mu}\Pi^{\mu 0}_{FD}({\bf q},{\bf q};\omega) = 
 \int d^3x\, \sum_{n<{\rm F}}
 \overline{U}_{n}({\bf x})\gamma^{0}{U}_{n}({\bf x}) =
 \int d^3x\, \rho_{\rm B}({\bf x}) 
 \equiv A \;.
 \label{CCPiFD2}
\end{equation}
In Fig.~\ref{Figure2} we display the cumulative violation of the 
vector current as a function of the angular-momentum channel 
$J^{\pi}$. Note that the plot also includes the corresponding 
violation in the Density-Feynman part of the nuclear polarization 
which is given by
\begin{equation}
 q_{\mu}\Pi^{\mu\beta}_{DF}({\bf q},{\bf q}';\omega) = -
 \rho^{\beta}({\bf q}\!-\!{\bf q}') \;.
\label{CCPiDF}
\end{equation}
In this way current conservation is properly restored:
\begin{equation}
 q_{\mu}\Pi^{\mu\beta}_{D}({\bf q},{\bf q}';\omega) = 
 q_{\mu}\left[\Pi^{\mu\beta}_{FD}({\bf q},{\bf q}';\omega)  +
         \Pi^{\mu\beta}_{DF}({\bf q},{\bf q}';\omega)\right] = 0 \;.
\label{CCPiD}
\end{equation}
Note that current conservation is maintained for each individual
$J^{\pi}$-channel. Figure~\ref{Figure3} validates this statement 
by displaying the timelike component of the polarization alongside 
the longitude component ($\hat{3}\!=\!\hat{\bf q}$) for the
isoscalar-dipole state in ${}^{40}$Ca. These results emerge from 
two powerful demands. First, the interaction driving the nucleon 
propagator must be identical to the one generating the mean-field 
ground state. Second, the negative-energy part of the spectrum must 
be kept, otherwise the nucleon propagator fails to become the 
Green's function for the relevant Dirac problem. One of the 
great virtues of the nonspectral approach is that the 
negative-energy states are included automatically.

So far our discussion of current conservation has been limited to the
lowest-order polarization. Nevertheless, the conservation of the
vector current at the RPA level places no additional demands on the
formalism. Indeed, it relies exclusively on the conservation of the
vector current at the Hartree level and it is independent of the
nature of the residual interaction. This result may be derived from 
the structure of Dyson's equation for the nuclear polarization.
Using Eqs.~(\ref{PiabRPA}) and~(\ref{CCPiD}) we obtain
\begin{eqnarray}
  q_{\mu}\Pi^{\mu\beta}_{\rm RPA}({\bf q},{\bf q}';\omega) &=&
  q_{\mu}\Pi^{\mu\beta}_{D}({\bf q},{\bf q}';\omega) \nonumber \\
  &+& \int\!\frac{d^3k}{(2\pi)^{3}}
       \frac{d^3k'}{(2\pi)^{3}}
  \left[q_{\mu}\Pi^{\mu\lambda}_{D}({\bf q},{\bf k};\omega)\right]           
   V_{\lambda\sigma}({\bf k},{\bf k}';\omega)
  \Pi^{\sigma\beta}_{\rm RPA}({\bf k}',{\bf q}';\omega) = 0\;.
 \label{CCPiRPA}
\end{eqnarray}
We close this section with a brief comment. As the conservation of the
vector current is exact in our formalism, we are entitled to a minor
simplification: the longitudinal component of the current can be
systematically eliminated in favor of the timelike component. Thus,
the RPA equations may be reduced from a $5\!\times\!5$ to a
$4\!\times\!4$ set of integral equations by simply adopting a
modified longitudinal propagator of the form:
\begin{equation}
  D_{0}({\bf k},\omega) \equiv 
  \left(\frac{k_{\mu}^{2}}{{\bf k}^{2}} \right)D({\bf k},\omega) \;;
  \quad k_{\mu}^{2}\!=\!(\omega^{2}\!-\!{\bf k}^{2}) \;.
 \label{LongProp}
\end{equation}
Note that the  gauge component of the vector propagator 
[the $k_{\alpha}k_{\beta}$ term in Eq.~(\ref{vecprop})]
has been eliminated from any further discussion because 
the vector mesons do indeed couple to a conserved vector 
current.

\subsection{Spurious strength in the isoscalar-dipole response}
\label{spurious}
\begin{figure}[h]
\vskip0.25in
\leavevmode\centering\psfig{file=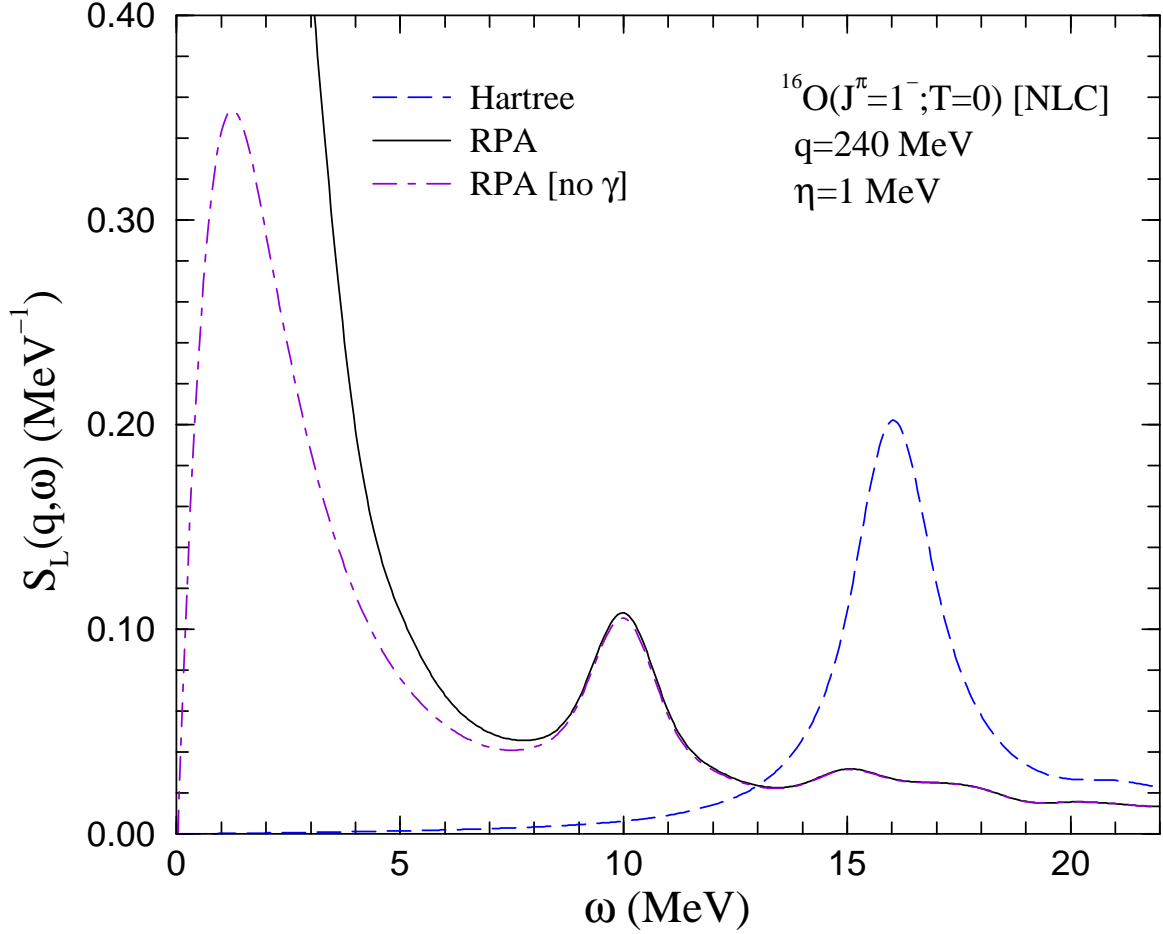,width=6in}
 \caption{Isoscalar dipole strength in ${}^{16}$O in
	  lowest-order Hartree (dashed line) and
	  in a consistent RPA (solid line) approximation. 
	  The dot-dashed line is the RPA response with
	  a residual interaction that lacks the
	  contribution from the isoscalar component of
	  the photon. The nonlinear model NLC was
	  employed in the calculation.}
 \label{Figure4}
\end{figure}

While we have argued earlier that the conservation of the vector
current at the RPA level is maintained irrespective of the nature 
of the residual particle-hole interaction, a consistent residual 
interaction becomes of utmost importance in the elimination of 
the spurious strength from the isoscalar-dipole response. This 
result, first demonstrated by Thouless for the nonrelativistic 
case~\cite{Th61} and later extended by Dawson and Furnstahl to 
the relativistic domain~\cite{DF90}, reinforces the importance 
of consistency in the formalism. As in the case of the conservation 
of the vector current, the decoupling of the spurious component
of the isoscalar-dipole response depends on the consistency 
between the residual particle-hole interaction and the 
particle-particle interaction driving the mean-field ground state. 
Figure~\ref{Figure4}, where the distribution of isoscalar-dipole 
strength in ${}^{16}$O is displayed, elucidates this point in a 
particularly clear fashion. The lowest-order Hartree response 
(dashed line) concentrates most of the isoscalar-dipole strength 
in a single fragment located around $\omega\!=\!16$~MeV of excitation 
energy. This is the region where many single-particle transitions 
from the p-shell to the sd-shell occur. Yet most of this strength 
is spurious, as evinced by the large amount being shifted 
to zero excitation energy in the RPA response (solid line). What 
remains is a relatively small fragment centered around 
$\omega\!=\!10$~MeV of excitation energy; we identify this 
fragment as the first physical isoscalar-dipole state in
${}^{16}$O. We have also included in Fig.~\ref{Figure4} an RPA
calculation (dot-dashed line) with a slightly ``tampered'' residual
interaction, namely, one that neglects the contribution from the
isoscalar component of the photon. Although much weaker than
its purely isoscalar (sigma and omega) counterparts, the photon
contribution remains indispensable at low-excitation energies. Indeed,
without it the spurious center-of-mass state fails to move all the 
way down to zero excitation energy.

\begin{figure}[h]
\vskip0.25in
\leavevmode\centering\psfig{file=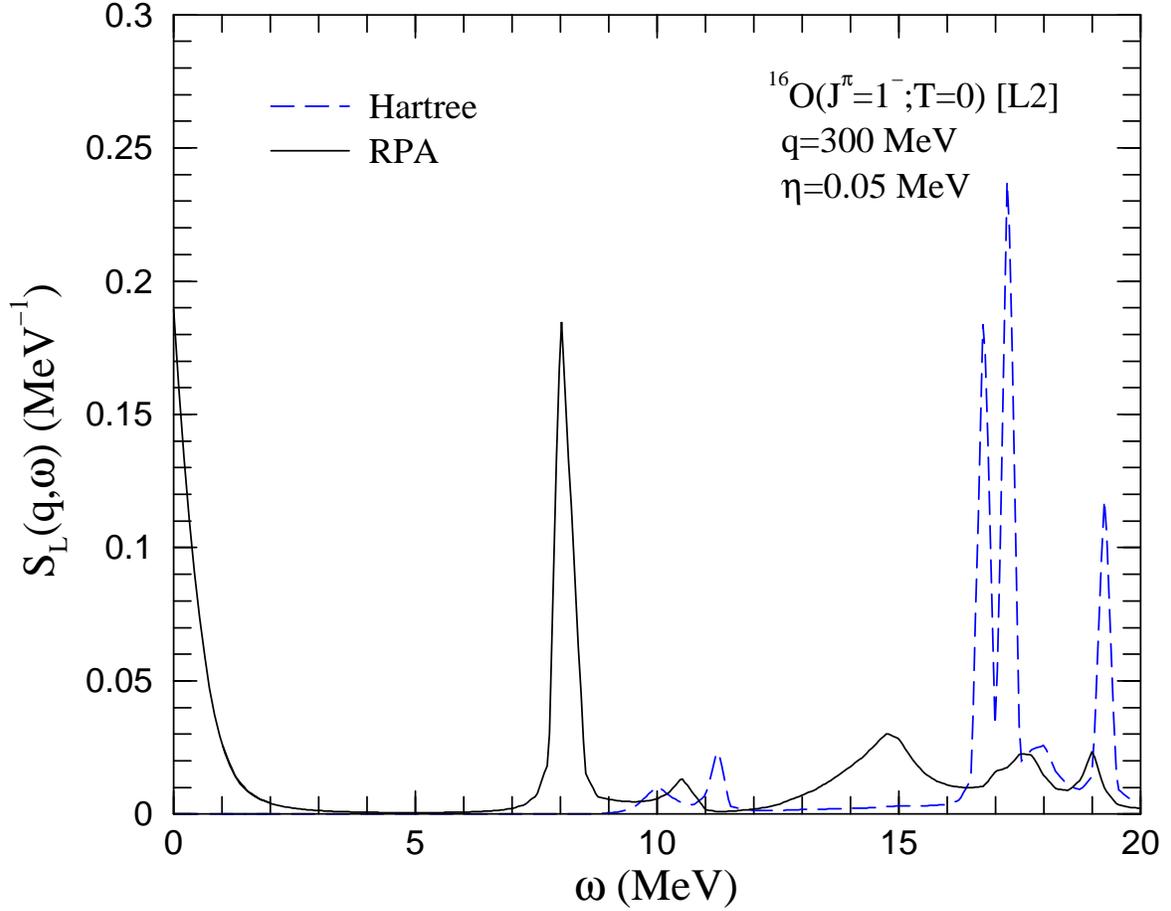,width=6in}
 \caption{Distribution of isoscalar-dipole strength 
	  in ${}^{16}$O in a lowest-order Hartree 
	  (dashed line) and in a consistent RPA 
	  (solid line) approximation. The linear
	  model L2 was employed in the calculation.}
 \label{Figure5}
\end{figure}

\begin{figure}[h]
\vskip0.25in
\leavevmode\centering\psfig{file=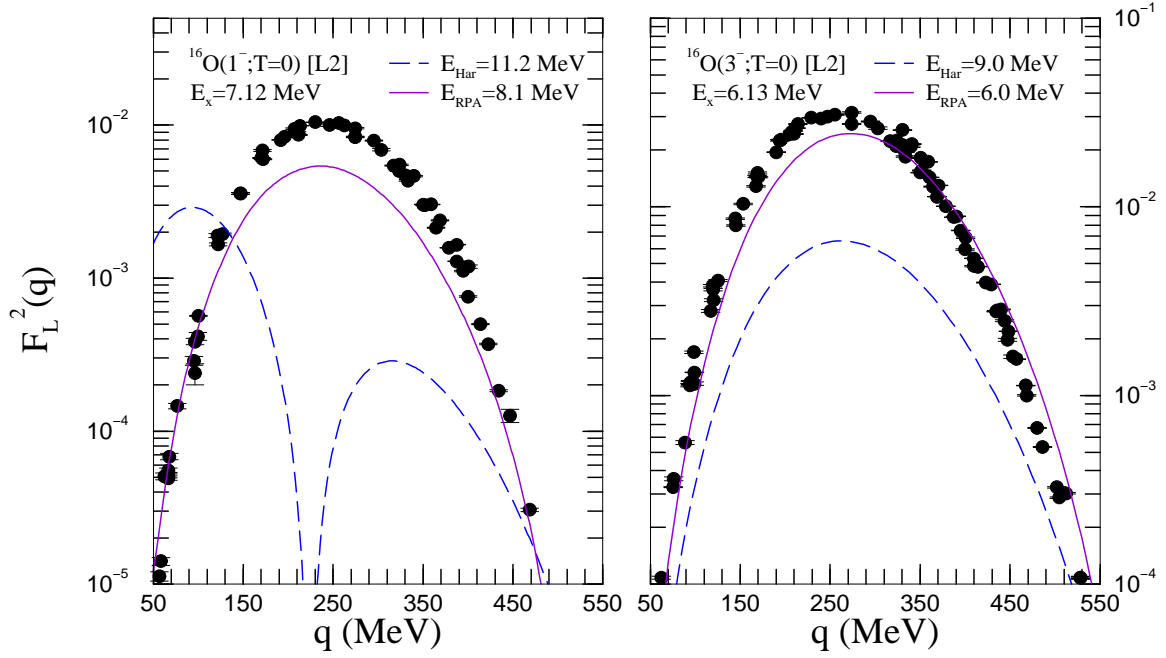,width=6in}
 \caption{Inelastic isoscalar-dipole (left panel) and 
	  isoscalar-octupole (right panel) form-factors 
	  for ${}^{16}$O in a lowest-order Hartree 
	  (dashed line) and in a consistent RPA (solid line) 
	  approximation. The linear model was employed in 
	  the calculation and the experimental data is from 
	  Ref.~\protect\cite{Bu86}.}
 \label{Figure6}
\end{figure}

A similar calculation for the linear L2-set is displayed in
Fig.~\ref{Figure5}. This time, however, the width has been reduced
considerably (from $\eta\!=\!1$~MeV to $\eta\!=\!0.05$~MeV) so that 
the various discrete single-particle excitations (dashed line) may 
be resolved. For example, the two small fragments in the 10-12 MeV 
region (dashed line) represent the proton and neutron 
$1P^{1/2}\rightarrow 2S^{1/2}$ single-particle excitations
respectively (see Table~\ref{Table2}). Moreover, by reducing
the width one removes any contamination from the spurious state 
into the first physical excitation (solid line). This is essential 
for a reliable extraction of the inelastic form-factor, which is
proportional to the area under the peak:
\begin{equation}
  F_{L}^{2}(q)=\lim_{\eta\rightarrow 0}
	       \frac{1}{4\pi}
	       \int_{\omega_{n}-\eta}^{\omega_{n}+\eta}
               S_{L}(q,\omega) d\omega \;.
 \label{FormFactor}
\end{equation}
Here $\omega_{n}$ represents the (discrete) excitation energy.  In
Fig.~\ref{Figure6} we show the isoscalar dipole form-factor extracted
from the longitudinal response. As we compare with actual experimental
data~\cite{Bu86}, the single-nucleon form-factor has been folded into
the calculation. The Hartree form-factor is the Fourier transform of
the $1P^{1/2}\rightarrow 2S^{1/2}$ single-particle transition density.
As such, it displays a very deep minimum due to the presence of a node
in the $2S^{1/2}$ wave function. Clearly, even a small amount of
configuration mixing will fill in this minimum.  Indeed, not only does
the RPA form-factor (solid line) shows no evidence of a minimum, but
it actually peaks very close to the Hartree minimum. Further, if the
separation between the spurious state and the physical states is
complete, then the momentum-transfer dependence of the isoscalar
dipole form-factor should display an octupole ($J\!=\!3$) behavior
rather than that of a dipole~\cite{DF90,Pi90,Sh89}. It may be seen in
Fig.~\ref{Figure6} that the $q$-dependence of the physical form-factor
is indeed (practically) identical to that of the octupole form-factor.

\section{Results}
\label{Results}
Having established the theoretical framework for the calculations
of the response, we now proceed to display our results for the 
distribution of isoscalar monopole and isoscalar dipole strength 
on a variety of closed-shell nuclei. As both monopole and dipole 
states can be excited through the timelike component of the vector 
current, we limit our discussion to the longitudinal response:
\begin{eqnarray}
  S_{\rm L}({\bf q},\omega) &=&
  \sum_{n}\Big|\langle\Psi_{n}|\hat{\rho}({\bf q})|
  \Psi_{0}\rangle\Big|^{2}
  \delta(\omega-\omega_{n}) \nonumber \\ &=&
  -\frac{1}{\pi}{\cal I}_m \Pi^{00}({\bf q},{\bf q},\omega) \;,
 \label{slong}
\end{eqnarray}
where $\hat{\rho}({\bf q})$ is the Fourier transform of the isoscalar
vector density, $\Psi_{0}$ is the exact nuclear ground state, and
$\Psi_{n}$ is an excited state with excitation energy $\omega_{n}$. 
\subsection{Isoscalar Giant Monopole Resonance}
\label{monopole}
\begin{figure}[h]
\vskip0.25in
\leavevmode\centering\psfig{file=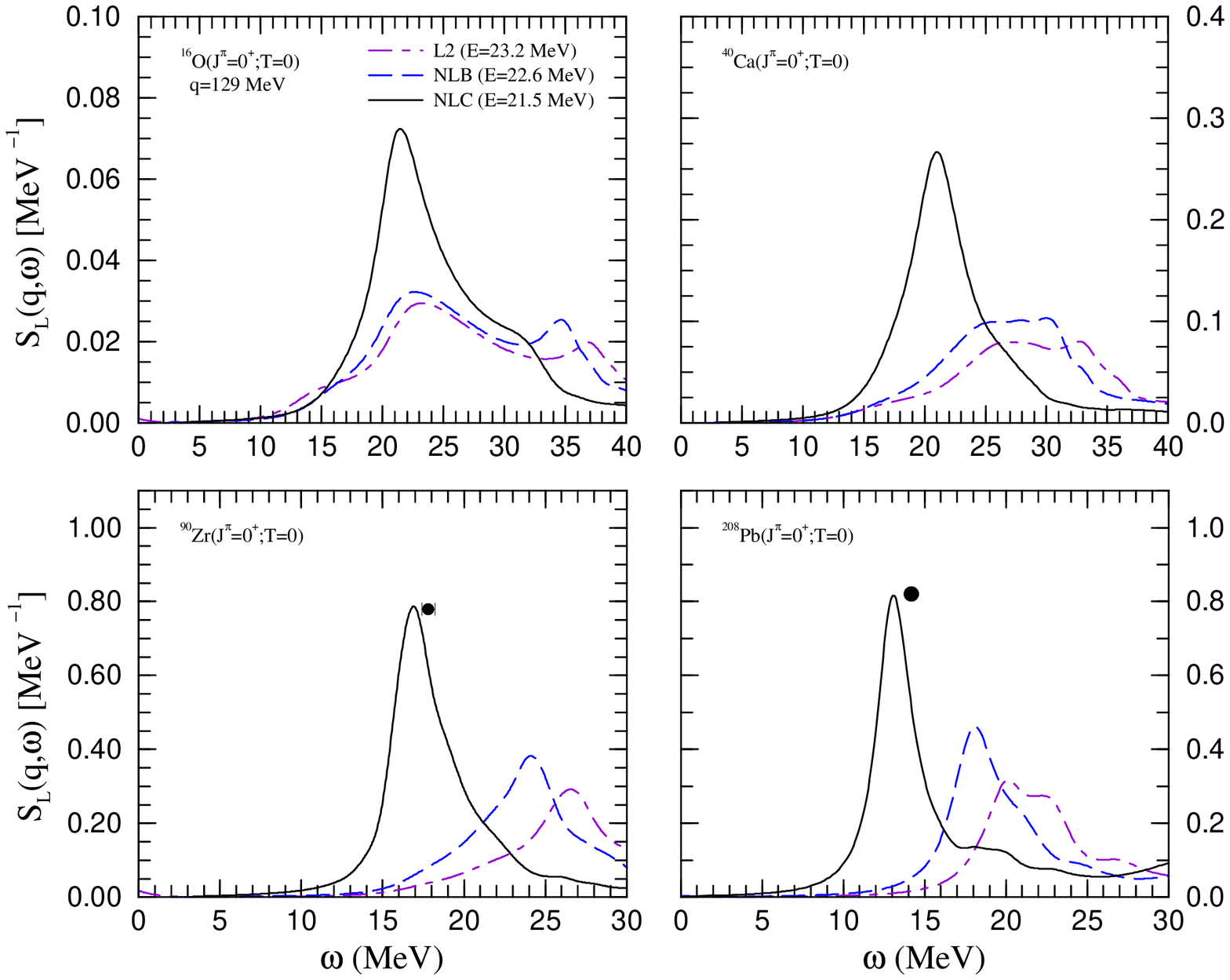,width=6in}
 \caption{Nuclear dependence of the isoscalar giant monopole 
          resonance in three relativistic mean-field models.}
 \label{Figure7}
\end{figure}

The isoscalar giant monopole resonance is the quintessential
compressional mode. Regarded as the ``breathing mode'' of the nucleus,
this excitation holds a special place in nuclear physics as it
provides, perhaps more than any other measurable observable, the most
direct determination of the compressibility of nuclear matter.  In
Fig.~\ref{Figure7} we display the distribution of isoscalar monopole
strength for various closed-shell nuclei as predicted by three
relativistic mean-field models. These predictions are also compiled 
in Table~\ref{Table3}. The three models have been defined in
Ref.~\cite{SW97} as L2, NLB, and NLC and have been constrained to
reproduce several bulk properties of nuclear matter at saturation as
well as the root-mean-square charge radius of ${}^{40}$Ca; the last
two models include self-interactions among the scalar field. The model
parameters have been listed in Table~\ref{Table1}. First discovered in
$\alpha$-scattering experiments from ${}^{208}$Pb~\cite{Yo77} and
recently measured with better precision, the peak of the GMR has been
reported to be located at an excitation energy of
$E\!=\!14.2\!\pm\!0.1$~\cite{Yo99}. As reported in a recent
publication~\cite{Pi00}, we found reasonable agreement between
experiment and our theoretical calculations using set NLC. The other
two sets, with compression moduli larger than $K\!=\!420$~MeV, predict
the location of the GMR at too large an excitation energy. This
behavior continues all through the periodic table.  Indeed, for
medium-size nuclei, such as ${}^{16}$O and ${}^{40}$Ca, it becomes
difficult to even identify a genuine GMR with parameter sets L2 and
NLB. In contrast, the identification of the GMR with parameter set NLC
is unambiguous for all nuclei and its prediction for the location of
the GMR in ${}^{90}$Zr is in good agreement with experiment. Finally,
acceptable agreement has been found with empirical formulas that
suggest that the position of the GMR should scale as the square root
of the compressibility. For example, peak energies for this mode have
been computed in the ratio of 1:1.38:1.53 for ${}^{208}$Pb and
1:1.43:1.57 for ${}^{90}$Zr, while the square root of the
nuclear-matter compressibilities are in the ratio of
1:1.37:1.56. These results suggest that models of nuclear structure
having compression moduli well above $K\!\approx\!200$~MeV are likely
to be in conflict with experiment.

\subsection{Isoscalar Giant Dipole Resonance}
\label{dipole}
The special role played by the isoscalar giant dipole resonance in
constraining theoretical models of the nuclear response has been
discussed extensively in previous sections. We trust that our results
have convinced the reader that the approach is sound and that the
spurious contamination has been efficiently removed from the physical
excitations. Hence, in the remainder of this section we focus on the
nuclear and model dependence of the ISGDR.  Moreover, we also discuss 
the substantial amount of isoscalar dipole strength predicted to exist 
at low energy and already observed experimentally. 

\begin{figure}[h]
\vskip0.25in
\leavevmode\centering\psfig{file=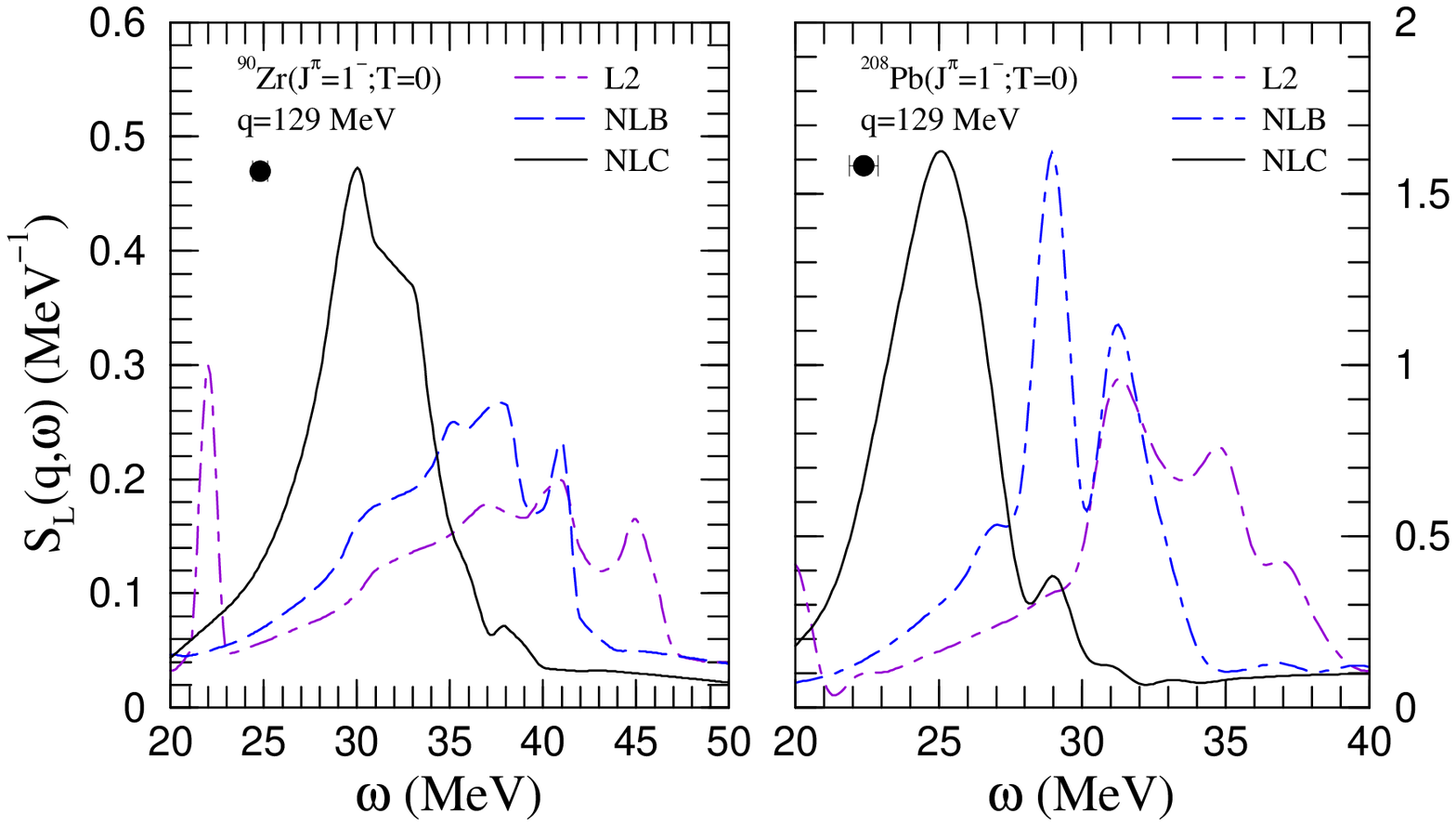,width=6in}
 \caption{Nuclear dependence of the isoscalar giant dipole
          resonance in three relativistic mean-field models.}
 \label{Figure8}
\end{figure}

The distribution of isoscalar-dipole strength in ${}^{90}$Zr and
${}^{208}$Pb is displayed in Fig.~\ref{Figure8} for the three
relativistic models. Note that no attempt has been made to identify a
resonant peak for the case of ${}^{16}$O and ${}^{40}$Ca as the
strength becomes too fragmented. As remarked earlier, the model with
the softest equation of state (NLC) provides the best description of
the experimental data~\cite{Pi00}. Thus, as hoped, the high-energy
component of the isoscalar-dipole response provides an independent
determination of the compression modulus of nuclear matter. Moreover,
it constraints, more than any other observable, theoretical models of
the nuclear response. Even so, we should note that the most accurate
of the models (NLC) still overpredicts by almost 5~MeV the energy of
the isoscalar dipole mode in ${}^{90}$Zr.

\begin{figure}[h]
\vskip0.25in
\leavevmode\centering\psfig{file=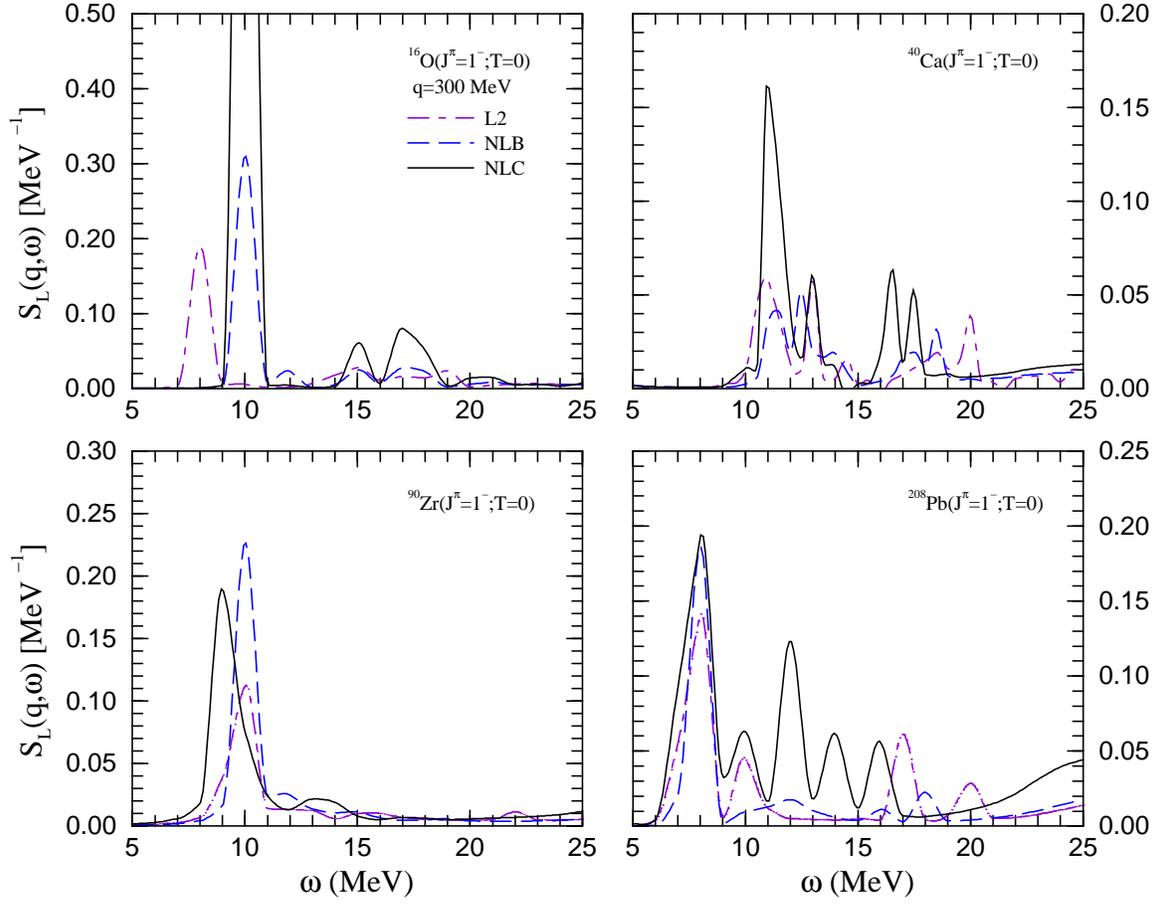,width=6in}
 \caption{Low-energy component of the isoscalar dipole
          strength in three relativistic mean-field models.}
 \label{Figure9}
\end{figure}

\begin{figure}[h]
\vskip0.25in
\leavevmode\centering\psfig{file=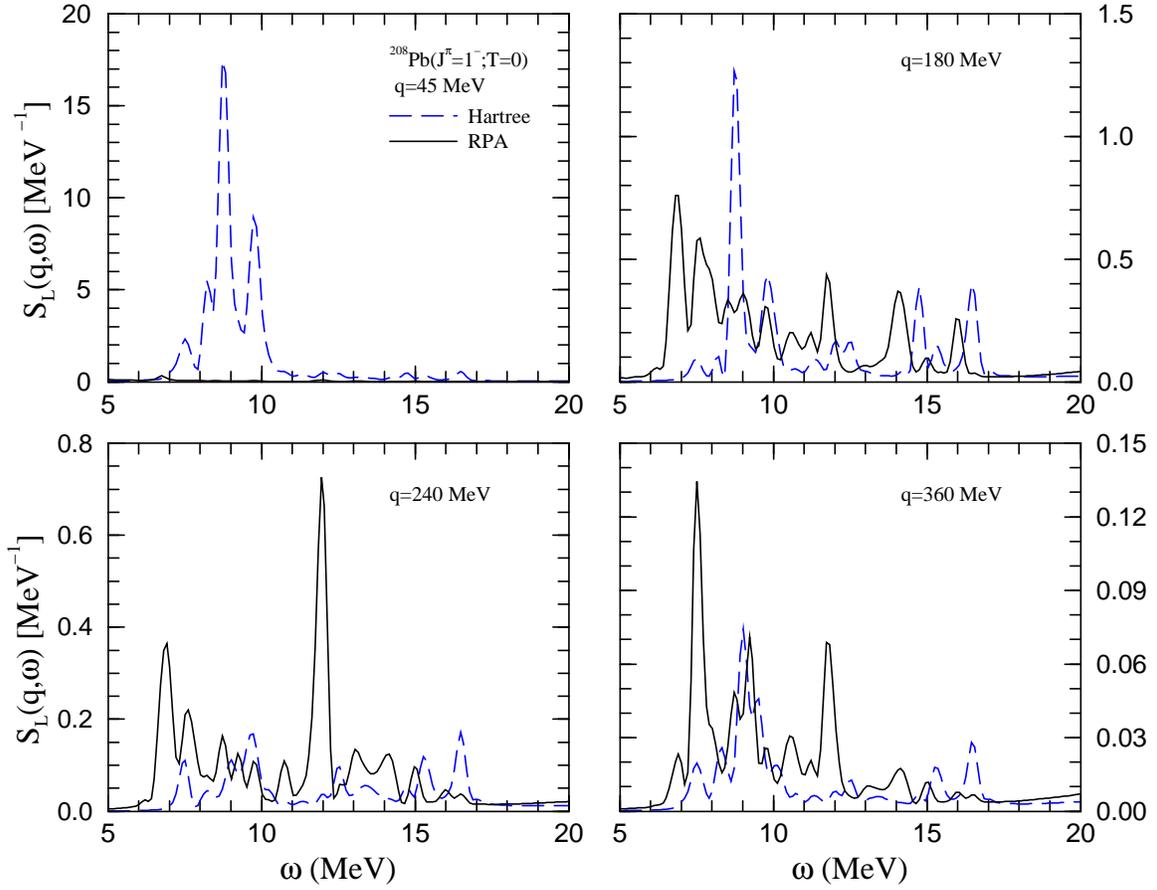,width=6in}
 \caption{Momentum-transfer dependence of the low-energy 
	  component of the isoscalar dipole strength in 
	  a lowest-order Hartree (dashed line) and RPA 
	  (solid line) approximations. All calculations
	  were performed using the NLC set.}
 \label{Figure10}
\end{figure}

In contrast to the high-energy component of the isoscalar-dipole
response, the low-energy component is independent of the compression
modulus of nuclear matter (see Fig.~\ref{Figure9}). Indeed, the
lowest-energy fragment in ${}^{208}$Pb is located at an excitation
energy of about 8 MeV --- irrespective of the parameters of the
model. That is, relativistic models having compression moduli ranging
from 220 MeV all the way up to 550 MeV predict a similar distribution
of low-energy isoscalar-dipole strength in ${}^{208}$Pb. This behavior
continues all throughout the periodic table. While the
extraction of a sole RPA state, and thus of an associated form-factor,
is difficult in the case of heavy nuclei, some interesting features
emerge from the study of the momentum-transfer dependence of the
distribution of strength. Figure~\ref{Figure10} displays such a
dependence for ${}^{208}$Pb. It shows that the large amount of
spurious strength observed at low-momentum transfer ($q\!=\!45$~MeV)
in the Hartree response gets shifted to zero excitation energy (not
shown in this figure) leaving a barely visible physical fragment at
around 7-8 MeV. Moreover, the evolution of RPA strength with momentum
transfer seem to follow the trends displayed by the inelastic form
factor of ${}^{16}$O (see Fig.~\ref{Figure6}). It has been proposed
in Ref.~\cite{Vr00}, from an analysis of the velocity fields, that 
the low-energy component of the isoscalar dipole mode is determined 
by surface effects.

\section{Conclusions}
\label{Conclusions}

The distribution of isoscalar monopole and isoscalar dipole strength 
has been computed in a relativistic random-phase-approximation to the
Walecka model using various parametrizations that incorporate scalar
self-interactions. While all of these models provide an equally good
description of the properties of nuclear matter at saturation, their
predictions for the nuclear compressibility differ by more than a
factor of two. Predictions for the energy of various surface modes in
medium-mass nuclei within a self-consistent random-phase-approximation
to the Walecka model have existed for over a decade. However, attempts
at calculating compressional modes in the original model, with a
compressibility of $K\!=\!545$ MeV, were doomed to failure.  Recently,
however, scalar self-interactions, so instrumental in softening the
equation of state, were incorporated into the calculation of the
response. Unfortunately, as lessons were being learned, others were
being forgotten. Chief among these was the important role of the
negative-energy states in the formalism.

In this paper the relativistic RPA formalism with scalar
self-interactions has been reviewed in great detail. A nonspectral
approach has been implemented that automatically includes both
positive- and negative-energy continuua without any reliance on
artificial cut-offs and truncations. Special emphasis was placed on
the role of self-consistency which demands that the same interaction
used to generate the mean-field ground state be used to: (i) compute
the nucleon propagator and (ii) the RPA response.  Enforcing (i)
guarantees the conservation of the vector current, while enforcing
(ii) successfully decouples the spurious isoscalar-dipole strength
from the physical response. A novel relation that quantifies the
violation of the vector current exclusively in terms of ground-state
form-factors was introduced. This relation may be used as a stringent
test on the numerics.

Predictions for the isoscalar giant-monopole resonance in the NLC
model, with a nuclear compressibility of $K\!=\!224$~MeV, were in good
agreement with experiment and also with semi-empirical formulas that
suggest that the position of the GMR should scale as the square root
of the nuclear compressibility. For the isoscalar-dipole mode the best
description of the data was still obtained with the NLC set, but here
the discrepancies were larger than in the monopole case.  In
particular, theoretical calculation overestimate the position of the
ISGDR in ${}^{90}$Zr by almost 5~MeV. In addition to the high-energy
component of the ISGDR, a low-energy component that is insensitive to
the compressibility of the model was clearly identified in all nuclei.
It has been proposed elsewhere that the low-energy component of the
isoscalar dipole mode is determined by surface effects. The existing
discrepancies between theory and experiment, particularly in the case 
of the ISGDR in ${}^{90}$Zr, are significant. The resolution of this
differences demands substantial effort on both fronts. While
extracting moments of the distribution will continue to be useful, we
suggest that in future studies the full distribution of strength be
adopted for comparisons between theory and experiment.
\appendix
\section{The $J^{\pi}$ content of the nuclear polarization}
\label{appendix}

The first step into the calculation of the RPA response is the
computation of the lowest-order polarization given in 
Eq.~(\ref{piqq}). Although this step apparently requires the 
evaluation of a six-dimensional integral, the spherical nature 
of the underlying mean-field potential enables one to carry out 
the four angular integrals analytically leaving a two-dimensional 
integral to be performed numerically. Thus, through a multipole 
decomposition, the density-dependent part of the timelike 
polarization may be written as~\cite{HP89,Pi90},
\begin{equation}
  \Pi^{00}_{D}({\bf q},{\bf q}';\omega) = \sum_{J=0}^{\infty}
  \Pi^{00}_{J}(q,q';\omega)
   P^J_{00}({\hat{\bf q}},{\hat{\bf q}'}) \;,
 \label{multipole}
\end{equation}
where all the dynamical information is contained in  
$\Pi^{00}_{J}(q,q';\omega)$ and the ``geometrical''
(or angular) dependence is given by the function
\begin{equation}
   P^J_{\lambda\lambda^\prime}({\hat{\bf q}},{\hat{\bf q}'})
   \equiv
   \sum_{M} D^{J }_{M\lambda}({\hat{\bf q}})
            D^{J*}_{M\lambda^\prime}({\hat{\bf q}'})\;. 
 \label{pfunction}
\end{equation}
Here $D^{J }_{M\lambda}({\hat{\bf q}})$ are the Wigner 
D-functions. Two of these functions may be combined by
using the following identity~\cite{HP89,Pi90}, 
\begin{equation}
   \int d{\hat{\bf k}}
   P^J_{\lambda\sigma}({\hat{\bf q}},{\hat{\bf k}})
   P^{J^\prime}_{\sigma\lambda^\prime}({\hat{\bf k}},{\hat{\bf q}'}) =
   {4\pi\over 2J+1} \delta_{JJ^\prime}
   P^{J}_{\lambda\lambda^\prime}({\hat{\bf q}},{\hat{\bf q}'}) \;,
\label{orthop}
\end{equation}
so that the three-dimensional integral equation required 
for the evaluation of the RPA polarization be reduced to 
a one-dimensional one, albeit one for each angular-momentum 
channel. Computing any specific multipole of the polarization 
insertion requires the evaluation of various reduced matrix 
elements, which are constrained by angular-momentum and parity 
selection rules. Because of the timelike nature of the vertex 
($\gamma^{0}$) only natural-parity states, such as the isoscalar 
monopole and dipole compressional modes, may be excited. 

Note that there are large computational demands imposed on an 
RPA calculation of a heavy nucleus. As the RPA equations 
Eq.~(\ref{PiabRPA}) are solved using standard matrix-inversion 
techniques~\cite{Pr88}, the lowest-order polarization must be 
computed on every point of a square momentum-transfer grid and 
for every polarization insertion that mixes with $\Pi^{00}_{D}$. 
The lowest-order polarization must therefore be evaluated several 
thousands times for a reliable extraction of the RPA response.
\acknowledgements
\medskip

This work was supported in part by the DOE under Contract 
No.DE-FG05-92ER40750.
%
%

\begin{table}
\caption{Various relativistic parameter 
         sets~\protect\cite{SW97}. The
         scalar mass and $\kappa$ are 
	 given in MeV.} 
 \begin{tabular}{crcccrr}
 Set & $g_{\rm s}^{2}$ & $g_{\rm v}^{2}$ 
     & $g_{\rho}^{2}$  & $m_{\rm s}$ 
     & $\kappa$ & $\lambda$ \\
 \hline 
 L2    & 109.63 & 190.43 & 65.23 & 520 &    0 &    0 \\
 NLB   &  94.01 & 158.48 & 73.00 & 510 &  800 &   10 \\
 NLC   &  95.11 & 148.93 & 74.99 & 501 & 5000 & -200 
  \label{Table1}
 \end{tabular}
\end{table}
%
\begin{table}
\caption{Bound single-particle orbitals in $^{16}$O
	 and low-energy dipole (single-particle) 
	 transitions in three different relativistic 
  	 models. All energies are given in  MeV.}
 \begin{tabular}{ccccccc}
 Orbital &  L2-n &  L2-p  
         & NLB-n & NLB-p 
         & NLC-n & NLC-p \\
 \hline 
 $1S^{1/2}$ & $41.39$ & $37.17$ 
            & $38.75$ & $34.59$ 
            & $39.33$ & $35.18$ \\
 $1P^{3/2}$ & $20.57$ & $16.68$ 
            & $19.89$ & $16.02$ 
            & $20.77$ & $16.91$ \\
 $1P^{1/2}$ & $12.53$ & $ 8.77$ 
            & $14.10$ & $10.30$ 
            & $15.46$ & $11.65$ \\
 $1D^{5/2}$ & $ 3.34$ & --- 
            & $ 3.44$ & --- 
            & $ 4.46$ & $ 1.03$  \\
 $2S^{1/2}$ & $ 1.35$ & ---  
            & $ 1.55$ & ---  
            & $ 2.50$ & ---      \\
 \hline 
 Transition & & & & \\
 \hline
 $1P^{1/2}\rightarrow2S^{1/2}$ 
            & $11.18$ & $\sim 10$ 
            & $12.50$ & $\sim 11$
            & $12.95$ & $\sim 12$ \\
 $1P^{3/2}\rightarrow1D^{5/2}$ 
            & $17.23$ & $\sim 17$
            & $16.45$ & $\sim 16$  
            & $16.31$ & $15.88$   \\
 $1P^{3/2}\rightarrow2S^{1/2}$ 
            & $19.22$ & $\sim 18$
            & $18.30$ & $\sim 17$
            & $18.27$ & $\sim 17$
  \label{Table2}
 \end{tabular}
\end{table}
%
\begin{table}
\caption{Nuclear dependence for the energy of the 
	 isoscalar giant monopole resonance in three 
	 different relativistic models. All energies 
	 are given in MeV.}
 \begin{tabular}{ccccc}
 Model & ${}^{16}$O 
       & ${}^{40}$Ca  
       & ${}^{90}$Zr     
       & ${}^{208}$Pb \\
 \hline 
 L2    & 23.2  & 27.3   &  26.5  &  20.1   \\
 NLB   & 22.6  & 27.9   &  24.1  &  18.1   \\  
 NLC   & 21.5  & 21.0   &  16.9  &  13.1   \\ 
 \hline
 Exp.  & --- & --- & $17.8 \pm 0.4$ & 14.2 $\pm$ 0.1
  \label{Table3}
 \end{tabular}
\end{table}
%

\end{document}